\documentclass[prd,twocolumn,showpacs,floatfix,amsmath,nofootinbib,amssymb,floatfix]{revtex4}
\usepackage{graphicx,color,dcolumn,booktabs,bm}
\usepackage{longtable,lscape}
\usepackage{pdfpages}
\usepackage{txfonts}
\usepackage{overpic}
\usepackage{amssymb}
\usepackage{indentfirst}
\usepackage{feynmf}   %{feynmp}
\usepackage{slashed}  %for Feynman symbols
\usepackage{cases}
\usepackage{color}
\usepackage{multirow}
\usepackage{threeparttable}
\usepackage{epstopdf}
\usepackage{enumerate}
\usepackage{graphicx,color,dcolumn,booktabs,bm}
\usepackage[colorlinks, citecolor=blue,anchorcolor=red,menucolor=red, linkcolor=red,filecolor=red,urlcolor=blue,frenchlinks=red]{hyperref}

\graphicspath{{Figures/}} %

\begin{document}
%\begin{CJK}{GBK}{}
%Newly observed $\Xi^*_b(6227)^-$ at LHCb as a first member in the $P$-wave $\Xi_b^\prime$ family
%Spectroscopy and strong decay of the low-excited bottomed-strange baryons in the nonrelativistic quark model

\title{Role of newly discovered $\Xi_b(6227)^-$ for constructing excited bottom baryon family}
\author{Bing Chen$^{1,3}$}\email{chenbing@shu.edu.cn}
\author{Ke-Wei Wei$^1$}\email{weikw@hotmail.com}
\author{Xiang Liu$^{2,3}$\footnote{Corresponding author}}\email{xiangliu@lzu.edu.cn}
\author{Ailin Zhang$^{4}$}\email{zhangal@staff.shu.edu.cn}
\affiliation{$^1$Department of Physics, Anyang Normal University,
Anyang 455000, China\\$^2$School of Physical Science and Technology,
Lanzhou University,
Lanzhou 730000, China\\
%urlcolor=blue
$^3$Research Center for Hadron and CSR Physics, Lanzhou University
$\&$ Institute of Modern Physics of CAS,
Lanzhou 730000, China\\
$^4$Department of Physics, Shanghai University, Shanghai 200444,
China}

\date{\today}

\begin{abstract}
Selecting the newly observed $\Xi_b(6227)^-$ by LHCb as a study example, we decode its inner structure by giving the mass spectrum analysis and the investigation of its two-body strong decay behaviors. Our result indicates that the $\Xi_b(6227)^-$ is a good candidate of the $P$-wave $\Xi_b^{\prime}$ state with $J^P=3/2^-$ or $5/2^-$. In addition, we further provide the information of the properties of the partners of the $\Xi_b(6227)^-$. These predicted states include three $2S$ states and the remaining $1P$ states in the bottom-strange baryon family. The calculated sizable Okubo-Zweig-Iizuka$-$allowed decay widths of these partners show that the experimental search for them becomes possible via LHCb. We have a reason to believe that the present study can be treated as a start point for constructing the highly excited bottom baryon spectroscopy.
\end{abstract}
 \maketitle

\section{Introduction}\label{sec1}

As an important part of the hadron spectrum, the heavy baryon family is being established step by step with the cooperative efforts from both experimentalists and theorists. A typical example is the charmed baryon group. In the past 20 years, about 20 charmed baryon candidates have been announced by the different experimental collaborations~\cite{Patrignani:2016xqp}. However, searching for the bottom baryon states is not an easy task for experiment since production of the bottom baryons seems to be more difficult than that of the charmed baryon case.

\begin{figure}[htbp]
\begin{center}
\includegraphics[width=8.6cm,keepaspectratio]{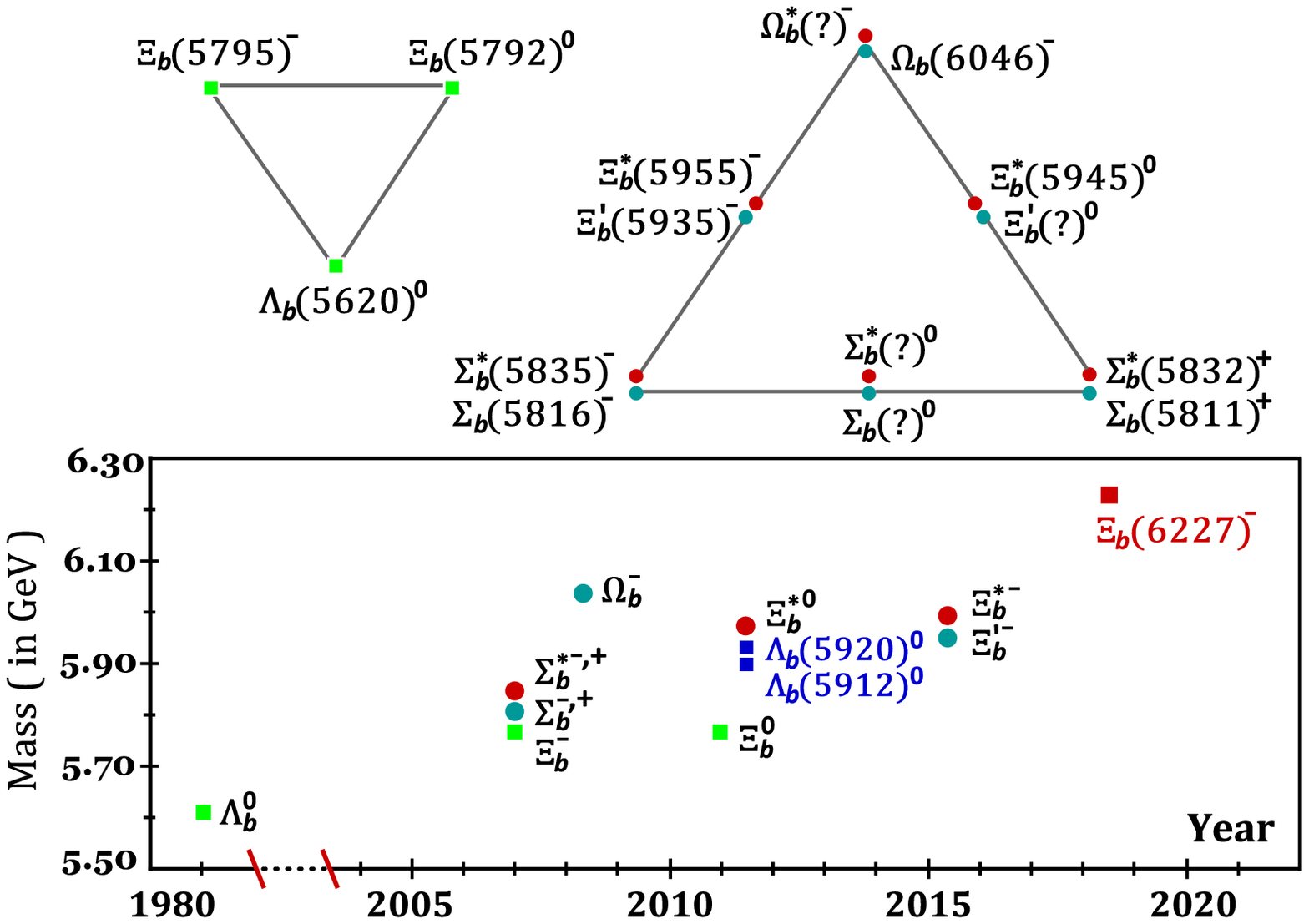}
\caption{The discovered bottomed baryons and the established $1S$ baryon multiplets.}\label{Fig1}
\end{center}
\end{figure}

The first discovered bottom baryon state, $\Lambda_b(5620)^0$, was reported by CERN R415 in 1981 \cite{Basile:1981wr}, while the first evidence of the $\Xi_b(5792)^0$ and the
$\Xi_b(5795)^-$ with $I(J^P)=1/2(1/2^+)$ was reported by DELPHI in 1995 \cite{Abreu:1995kt}. In the following 12 years, almost no experimental progress on searching for the new bottom baryons was made. This situation was changed in 2007 since the $\Sigma_b(5815)^-$ and $\Sigma^{\ast}_b(5835)^-$ with $I(J^P)=1(1/2^+)$ and $1(3/2^+)$ respectively were discovered by CDF \cite{Aaltonen:2007ar}. After that, more bottom baryons were reported, especially with the running of the LHCb \cite{Aaij:2012da,Aaij:2014yka,Aaij:2016jnn}.
We summarize all observed bottom baryons in Fig.~\ref{Fig1}. Until now, the $1S$ states of bottom baryon family almost have been established experimentally except the $\Sigma_b^0/\Xi_b^{\prime0}$ with $J^P=1/2^+$ and the $\Sigma_b^{*0}/\Omega_b^{\ast-}$ with $J^P=3/2^+$.
In addition, two $1P$ $\Lambda_b$ candidates, i.e., the $\Lambda_b(5912)^0$ and the $\Lambda_b(5920)^0$, have been observed by the LHCb~\cite{Aaij:2012da} and CDF~\cite{Aaltonen:2013tta}. Other excited bottom baryons are waiting for exploration. Obviously, it is time for the LHCb.

Very recently, the LHCb brought us a surprise since a new bottom baryon state, the $\Xi_b(6227)^-$, was found in both $\Lambda^0_bK^-$ and $\Xi^0_b\pi^-$ channels~\cite{Aaij:2018yqz}, which has the resonance parameters
\begin{equation}
\begin{split}
&M=6226.9\pm2.0\textrm{(stat)}\pm0.3\textrm{(syst)}\pm0.2(\Lambda_b^0)\textrm{~MeV},\\&
\Gamma=18.1\pm5.4\textrm{(stat)}\pm1.8\textrm{(syst)}\textrm{~MeV}.\nonumber
\end{split}
\end{equation}
We would like to emphasize that none of the excited bottom baryon states has been seen in its Okubo-Zweig-Iizuka (OZI)-allowed decay channels before the observation of the $\Xi_b(6227)^-$. Thus, the $\Xi_b(6227)^-$ is the first excited bottom baryon which was identified by the OZI-allowed decay modes. In addition, the measured relative production rates indicate that two allowed decay modes, i.e., $\Lambda_bK$ and $\Xi_b\pi$, may have nearly equal branching ratios for the $\Xi_b(6227)^-$ \cite{Aaij:2018yqz}.

Obviously, the $\Xi_b(6227)^-$ is a key state when we begin to establish whole excited bottom baryon spectroscopy. In this work, the main task is to reveal the inner structure of this newly observed state by the analysis of the mass spectrum and corresponding OZI-allowed decay behavior. We find that the $\Xi_b(6227)^-$ is most likely a $P$-wave $\Xi_b^{\prime}$ state with $J^P=3/2^-$ or $5/2^-$. Besides making such a conclusion, we also predict the properties of the partners of the $\Xi_b(6227)^-$, which include the $2S$ states and the remaining $1P$ states in the bottom-strange baryon sector. Since these states have the sizable OZI-allowed decay modes, experimental searching for them will be an interesting research issue. We believe that treating the $\Xi_b(6227)^-$ as a $P$-wave bottom baryon is not only a start point of constructing the whole excited bottom baryons but also can make the hadron spectroscopy become complete, which is helpful to further understand the nonperturbative behavior of QCD.

This paper is organized as follows. After the Introduction, we decode the newly reported $\Xi_b(6227)^-$ as a $P$-wave bottom-strange baryon in Sec. \ref{sec2}, where the analysis of mass spectrum and the calculation of strong decay behaviors can provide the direct support to this scenario. Finally, the paper ends with the discussion and conclusion in Sec. \ref{sec3}.

\section{Decoding the property of the $\Xi_b(6227)^-$}\label{sec2}

\iffalse
\begin{table*}[htbp]
\caption{The predicted masses of $1P$ and $2S$ bottomed-strange baryon states in MeV.} \label{table1}
\renewcommand\arraystretch{1.2}
\begin{tabular*}{156mm}{@{\extracolsep{\fill}}rcccccccccc}
\toprule[1pt]\toprule[1pt]
        & \multicolumn{2}{c}{$1P~(\Xi_b)$} &  $2S~(\Xi_b)$  & \multicolumn{5}{c}{$1P~(\Xi^\prime_b)$}  & \multicolumn{2}{c}{$2S~(\Xi^\prime_b)$} \\
\cline{2-3}\cline{4-4}\cline{5-9}\cline{10-11}
 $J^P$  & $1/2^-$  & $3/2^-$  & $1/2^+$  & $1/2^-$   & $1/2^{\prime-}$  & $3/2^-$   & $3/2^{\prime-}$  & $5/2^-$   & $1/2^+$   & $3/2^+$  \\
\midrule[0.8pt]
 This work                    & 6096     & 6102     & 6255      & 6249   &  6239    & 6244      & 6213   &  6217   & 6365   &  6376  \\
 Ref.~\cite{Roberts:2007ni}   & 6090     & 6093     & 6230      & 6305   &  6308    &           & 6190   &  6201   & 6349   &  6376  \\
 Ref.~\cite{Ebert:2011kk}     & 6120     & 6130     & 6266      & 6233   &  6227    & 6234      & 6224   &  6226   & 6329   &  6342  \\
 Ref.~\cite{Chen:2014nyo}     & 6097     & 6106     &           &        &          &           &        &         &        &        \\
 Ref.~\cite{Thakkar:2016dna}  & 6151     & 6141     & 6203      &        &          &           &        &         &        &        \\
 Ref.~\cite{Valcarce:2008dr}  & 6109     &          & 6258      & 6223   &          &           &        &         & 6360   &  6373  \\
\bottomrule[1pt]\bottomrule[1pt]
\end{tabular*}
\end{table*}
\fi

Since the $\Xi_b(6277)^-$ state was observed in the decay channels $\Lambda_b^0K^-$ and $\Xi_b^0\pi^-$, we conclude that it must contain $sbd$ quark component.
To reveal its property, a mass spectrum analysis should be given. In the past years,
the mass spectra of highly excited bottom baryons were investigated by several phenomenological models, which include the nonrelativistic quark model~\cite{Roberts:2007ni}, the QCD-motivated relativistic quark model~\cite{Ebert:2011kk}, the QCD motivated hypercentral quark model~\cite{Thakkar:2016dna}, the Faddeev formalism~\cite{SilvestreBrac:1996bg} or method~\cite{Valcarce:2008dr}, the relativistic flux tube model~\cite{Chen:2014nyo}, the QCD sum rule~\cite{Wang:2010it,Mao:2015gya}, and the Regge phenomenology~\cite{Wei:2016jyk}.
These studies of the mass spectrum of bottom baryon support the assignment of the $\Xi_b(6277)^-$ as a $2S$ or $1P$ state in the bottom-strange baryon sector.

As suggested in our former works~\cite{Chen:2016iyi,Chen:2017gnu}, the heavy-light baryon system can be treated as a quasi$-$two-body system in the heavy quark-light diquark picture.\footnote{The basic consideration of why we can simplify a heavy-light baryon system as a quasi-two-body
system in the heavy quark-light diquark picture and the details of calculation have been explained in Refs.~\cite{Chen:2016iyi,Chen:2017gnu}.} Thus, the Cornell potential~\cite{Eichten:1978tg} could be used to phenomenologically depict the confining interaction between a bottom quark and the light diquark.
Under this treatment, we may construct the following the Schr\"odinger equation:
\begin{equation}
\left(-\frac{\nabla^2}{2m_\mu}-\frac{4\alpha}{3r}+br+C+\frac{32\alpha\sigma^3}{9\sqrt{\pi}m_{di}m_b}\vec{\textrm{S}}_{di}\cdot\vec{\textrm{S}}_{b} \right)\psi_{nL} = E\psi_{nL}. \label{eq1}
\end{equation}
Here, $\vec{\textrm{S}}_{di}$ and $\vec{\textrm{S}}_{b}$ denote the spins of the light diquark and the $b$ quark, respectively. In Jaffe's terminology~\cite{Jaffe:2004ph}, the light scalar quark cluster in the $\Xi_b$ baryon system is called a ``good'' diquark ($S_{di}=0$), while the light axial-vector quark cluster in the $\Xi^\prime_b$ baryon is named the ``bad'' diquark ($S_{di}=1$). Therefore, the spin-spin contact hyperfine interaction in Eq. (\ref{eq1}) is important for calculating the masses of $nS$ $\Xi^\prime_b$ states. The reduced mass is defined as $m_\mu~{\equiv}~m_{di}m_b/(m_{di}+m_b)$. The parameters $\alpha$, $b$, and $C$ stand for the strength of the color Coulomb potential, the strength of linear confinement, and a mass-renormalized constant, respectively. By solving the Schr\"odinger equation, the spin-averaged masses of these excited bottom baryons can be obtained. When the spin-orbit and tensor interactions are included, all masses of $\lambda$-mode excited bottom-strange baryons can be calculated. All values of parameters used in our calculation are listed in Table \ref{table1}.

\begin{figure}[t]
\begin{center}
\includegraphics[width=8.6cm,keepaspectratio]{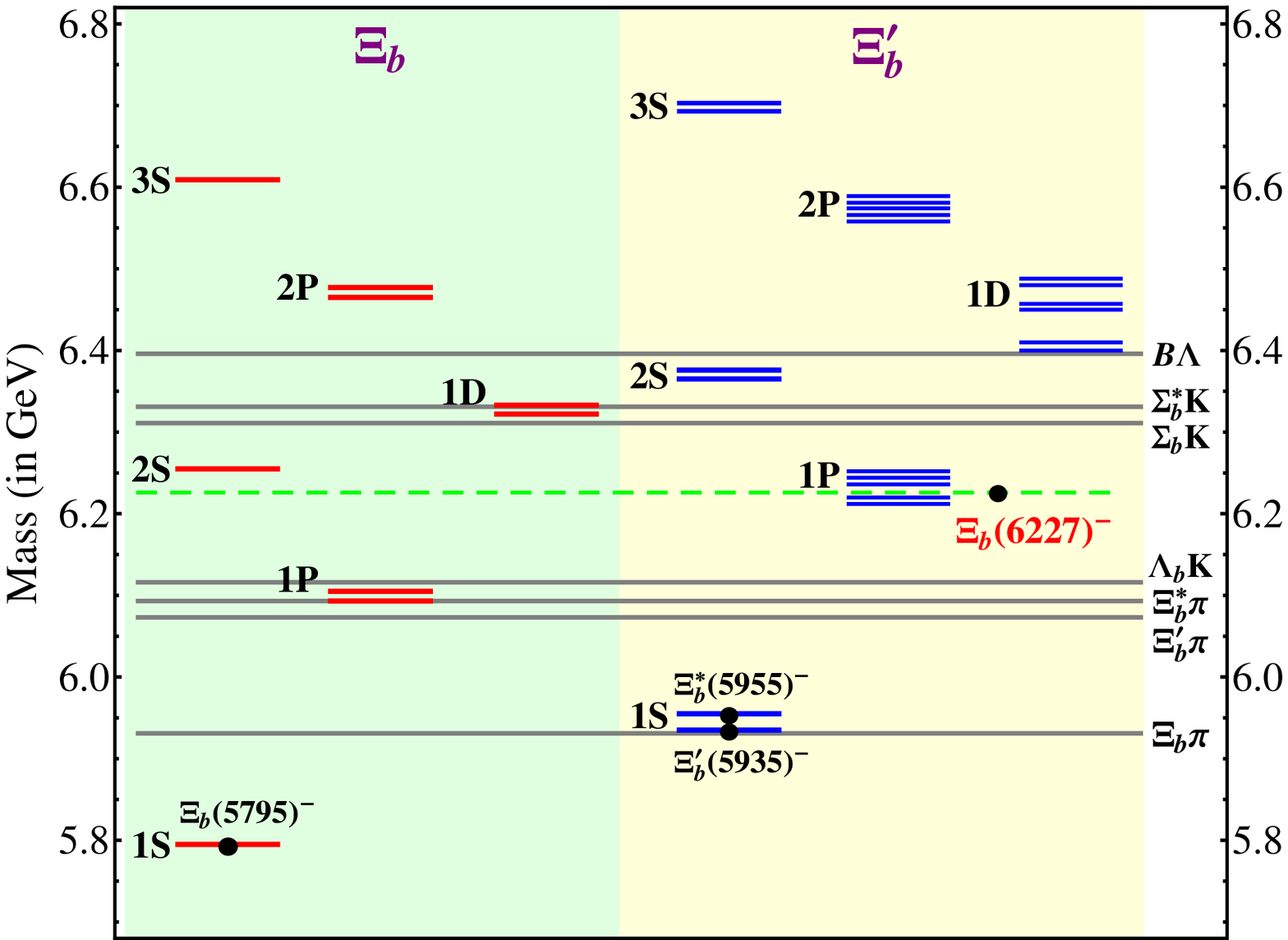}
\caption{The obtained masses for the bottom-strange baryons. The red solid lines (left) correspond to the predicted masses of $\Xi_b$ states which are composed of a good diquark and a bottom quark, while the blue solid lines (right) correspond to the $\Xi^\prime_b$ states which contain a bad diquark. Here, we also listed the measured masses of the ground states~\cite{Patrignani:2016xqp} and the $\Xi_b(6227)^-$~\cite{Aaij:2018yqz}, which are marked by ``filled circle''.}\label{Fig2}
\end{center}
\end{figure}

\begin{table}[htbp]
\caption{Values of the parameters for the bottomed baryons in the nonrelativistic quark potential model where the mass of $b$ quark is taken as 4.96 GeV. $m_{di}$ refers to the mass of different diquarks.
}\label{table1}
\renewcommand\arraystretch{1.3}
\begin{tabular*}{86mm}{c@{\extracolsep{\fill}}ccccc}
\toprule[1pt]\toprule[1pt]
 Parameters       & $m_{di}$ (GeV)    & $\alpha$       & $b$  (GeV$^2$)   & $\sigma$ (GeV)    & $C$ (GeV)  \\
\toprule[1pt]
$\Lambda_b$       & 0.45              & 0.20           & 0.112            & $-$               & 0.265  \\
$\Xi_b$           & 0.63              & 0.26           & 0.118            & $-$               & 0.176 \\
$\Sigma_b$        & 0.66              & 0.22           & 0.116            & 1.20              & 0.185  \\
$\Xi^{\prime}_b$  & 0.78              & 0.22           & 0.116            & 1.20              & 0.152  \\
$\Omega_b$        & 0.91              & 0.26           & 0.120            & 1.07              & 0.120  \\
\bottomrule[1pt]\bottomrule[1pt]
\end{tabular*}
\end{table}

We present the calculated masses of bottom-strange baryons in Fig. \ref{Fig2} and make a comparison of them with the experimental data. We also list several typical thresholds, which are denoted by the grey solid lines in Fig. \ref{Fig2}. Obviously, the quark potential model adopted here has reproduced the masses of three observed ground states of the bottom-strange baryon family. And then, we may find that the newly observed state, the $\Xi_b(6227)^-$, could be a candidate of a $2S$ $\Xi_b$ state or a $1P$ $\Xi^\prime_b$ state only by this mass spectrum analysis, where the theoretical value is close to the mass of the  $\Xi_b(6227)^-$ under the $2S$ and $1P$ assignments.
To give further constraints on its quantum number, we will investigate the strong decays in the following.

For carrying out the study of strong decay behaviors of the discussed bottom-strange baryons, we employ the quark pair creation (QPC) model~\cite{Micu:1968mk,LeYaouanc:1972vsx,LeYaouanc:1988fx} to calculate their two-body OZI-allowed decays. The QPC model has been extensively used to study the strong decays of different kinds of hadrons. %In our previous work~\cite{Chen:2016iyi}, the OZI-allowed decays of these observed charmed baryons are well described.
For a decay process of an excited $bsq$ baryon state ($q$ designates a $u$ or $d$ quark), $A(q(1)s(2)b(3))\rightarrow{B}(q(5)s(2)b(3))+C(q(1)\bar{q}(4))$, the transition matrix element in the QPC model is written as $\langle{BC}|\mathcal {\hat{T}}|A\rangle=\delta^3(\textbf{\textrm{K}}_B+\textbf{\textrm{K}}_C)\mathcal {M}^{j_A,j_B,j_C}(p)$, where the transition operator $\mathcal {\hat{T}}$ reads as
\begin{equation}
\begin{split}
\mathcal {\hat{T}}=&-3\gamma
\sum_{\text{\emph{m}}}\langle1,m;1,-m|0,0\rangle \iint
d^3\textbf{\textrm{k}}_4d^3\textbf{\textrm{k}}_5\delta^3(\textbf{\textrm{k}}_4+\textbf{\textrm{k}}_5)\\ &\times\mathcal
{Y}_1^m\left(\frac{\textbf{\textrm{k}}_4-\textbf{\textrm{k}}_5}{2}\right)\omega^{(4,5)}\varphi^{(4,5)}_0\chi^{(4,5)}_{1,-m}b^\dag_{4}(\textbf{\textrm{k}}_4)d^\dag_{5}(\textbf{\textrm{k}}_5)% \nonumber
\end{split}\label{eq2}
\end{equation}
in a nonrelativistic limit. Here, the $\omega_0^{(4,5)}$ and $\varphi^{(4,5)}_0$ are the color and flavor wave functions of the $\bar{q}_4q_5$ pair created from the vacuum, respectively. Therefore, $\omega^{(4,5)}=(R\bar{R}+G\bar{G}+B\bar{B})/\sqrt{3}$ and $\varphi^{(4,5)}_0=(u\bar{u}+d\bar{d}+s\bar{s})/\sqrt{3}$ are color and flavor singlets. The $\chi^{(4,5)}_{1,-m}$ represents the pair production in a spin
triplet state. The solid harmonic polynomial $\mathcal {Y}_1^m(\textbf{\textrm{k}})\equiv|\textbf{\textrm{k}}|\mathcal {Y}_1^m(\theta_k,\phi_k)$ reflects the momentum-space distribution of the $\bar{q}_4q_5$. The dimensionless parameter $\gamma$ describes the strength of the quark-antiquark pair created from the vacuum.

The useful partial wave amplitude is related to the helicity amplitude $\mathcal {M}^{j_A,j_B,j_C}(p)$ by
\begin{equation}
\begin{aligned}\label{eq3}
\mathcal {M}^{A\rightarrow B+C}_{LS}(p)=\frac{\sqrt{2L+1}}{2J_A+1}&\sum_{\text{$j_B$,$j_C$}}\langle L0Jj_A|J_Aj_A\rangle\\
&\times\langle J_Bj_B,J_Cj_C|Jj_A\rangle\mathcal
{M}^{j_A,j_B,j_C}(p),
\end{aligned}
\end{equation}
where the $J_i$ and $j_i$ ($i=$ $A$, $B$, and $C$) denote the total angular momentum and their projection of initial and final hadron states, respectively. $L$ denotes the orbital angular momenta between the final states $B$ and $C$. Finally, the partial width of $A\rightarrow BC$ is written in terms of the partial wave amplitudes as
\begin{equation}
\begin{aligned}\label{eq4}
\Gamma(A\rightarrow BC)=2\pi\frac{E_BE_C}{M_A}p\sum_{L,S}|\mathcal
{M}_{LS}(p)|^2
\end{aligned}
\end{equation}
in the $A$ rest frame. To obtain the concrete expressions of $\mathcal {M}^{j_A,j_B,j_C}(p)$, an integral $I^{l_A,m}_{l_B,l_C}(p)$ should be performed, which describes the overlap of the spatial functions of the initial state $(A)$, the created pair from the vacuum, and two final states ($B$ and $C$). Usually, the simple harmonic oscillator (SHO) wave function $\psi^m_{n_rL}(\textbf{\textrm{k}})=\mathcal{R}_{n_rL}(\beta, \textbf{\textrm{k}})\mathcal {Y}_{n_rL}^m(\textbf{\textrm{k}})$ is taken to construct the spatial wave function of a hadron state. In this way, the analytical $I^{l_A,m}_{l_B,l_C}(p)$ can be extracted. In our calculations, all values of the SHO wave function scale (denoted as ``$\beta$''), which reflect the distances between the light diquark and the $b$ quark in the bottomed baryons, are obtained by reproducing the realistic root mean square radius via Eq.~(\ref{eq1}). The results are collected in Table \ref{table2}. The values of $\beta$s for the light diquark and other related hadrons are taken from our previous work~\cite{Chen:2016iyi}.  In this work, the value of $\gamma$ is taken as 1.296 since the measured widths of $2S$ and $1P$ states of charmed and charmed-strange baryons have been reproduced in Ref.~\cite{Chen:2016iyi}.
\begin{table}[htbp]
\caption{The effective $\beta$ values for the different bottomed baryon states (in GeV).
}\label{table2}
\renewcommand\arraystretch{1.22}
\begin{tabular*}{82mm}{c@{\extracolsep{\fill}}ccccc}
\toprule[1pt]\toprule[1pt]
States         & $\Lambda_b$       &   $\Xi_b$  & $\Sigma_b$  & $\Xi'_b$     & $\Omega_b$   \\
\midrule[0.8pt]
 $1S(1/2^+)$   &    0.288          & 0.341      & 0.345       &    0.367     &  0.404 \\
 $1S(3/2^+)$   &                   &            & 0.334       &    0.355     &  0.390 \\
 $2S$          &    0.157          & 0.181      & 0.181       &    0.191     &  0.206 \\
 $3S$          &    0.115          & 0.131      & 0.132       &    0.139     &  0.148  \\
 $1P$          &    0.198          & 0.227      & 0.228       &    0.241     &  0.258  \\
 $2P$          &    0.131          & 0.149      & 0.150       &    0.158     &  0.169 \\
 $1D$          &    0.157          & 0.178      & 0.179       &    0.189     &  0.201 \\
\bottomrule[1pt]\bottomrule[1pt]
\end{tabular*}
\end{table}

With the preparation above, we first test our method by calculating the partial widths of these observed $1S$ bottom baryon states. At present, only the $1S$ bottomed baryons have been reported by experiments in their OZI-allowed decay channels.
In Table~\ref{table3}, we make a comparison between theoretical and experimental results of partial decay widths, which shows that the measured decay behaviors of the $\Sigma_b(5815)^-$, $\Sigma_b(5835)^{*-}$, $\Xi^{\prime}_b(5935)^-$, and $\Xi^{*}_b(5955)^-$ can be well explained. Thus, the reliability of our method is tested due to this success, which makes us to apply it to studying the strong decay behaviors of these discussed $1P$ and $2S$ $bsq$ states.

\begin{table}[htb]
\caption{The calculated and measured decays of the $1S$ $\Sigma_b$ and $\Xi^{\prime}_b$ baryons (in MeV).}\label{table3}
\renewcommand\arraystretch{1.3}
\begin{tabular*}{85mm}{@{\extracolsep{\fill}}ccc}
\toprule[1pt]\toprule[1pt]
Decay channels                                             & Prediction       & Experiments~\cite{Patrignani:2016xqp}    \\
\midrule[0.8pt]
 $\Sigma_b(5815)^-\rightarrow\Lambda_b^0\pi^-$             & 5.12             & $4.9^{+3.3}_{-2.4}$  \\
 $\Sigma_b(5835)^{*-}\rightarrow\Lambda_b^0\pi^-$          & 9.13             & $7.5\pm2.3$  \\
 $\Xi^{\prime}_b(5935)^-\rightarrow\Xi_b\pi$               & 0.05             & $<~0.08$,~C.L.=95\% \\
 $\Xi^{*}_b(5955)^-\rightarrow\Xi_b\pi$                    & 1.09             & $1.65\pm0.33$  \\
\bottomrule[1pt]\bottomrule[1pt]
\end{tabular*}
\end{table}

\subsection{$2S$ states}

As shown in Fig. \ref{Fig2}, there are three $2S$ states of bottom-strange baryons, i.e., the $\Xi_b(6255)$, $\Xi^{\prime}_b(6365)$, and $\Xi^{\prime}_b(6376)$. Hereafter we use the predicted masses to distinguish these unobserved states. Their predicted decay properties are collected in Table \ref{table4}.

The predicted $\Xi_b(6255)$ cannot decay into $\Lambda_bK$ and $\Xi_b\pi$ since these decay modes are forbidden for the $2S$ $\Xi_b$ state due to the requirement of the heavy quark symmetry. Our result indicates that the $\Xi^{\prime}_b(5935)\pi$ and $\Xi^{*}_b(5955)\pi$ are the two dominant decay channels for the $2S$ $\Xi_b$ state. And then, its total decay width is 16.5 MeV. If comparing the above information with experimental data of the observed $\Xi_b(6227)^-$, we may fully exclude the possibility of the $\Xi_b(6227)^-$ as a $2S$ $\Xi_b$ state, which is composed of a good diquark and a bottom quark. If experimentally establishing the predicted $\Xi_b(6255)$, we suggest searching for it via the decay channels $\Xi^{\prime}_b(5935)\pi$ and $\Xi^{*}_b(5955)\pi$.

The predicted $\Xi^{\prime}_b(6365)$ and $\Xi^{\prime}_b(6376)$ are bottom-strange baryons with a bad diquark and a bottom quark. Their calculated total decay widths in Table \ref{table4} are far larger than the measured width of the $\Xi_b(6227)^-$. According to this point, we may conclude that the $\Xi_b(6227)^-$ does not favor the $2S$ $\Xi_b^\prime$ assignments.
Moreover, the predicted masses of two $2S$ $\Xi_b^\prime$ are higher than the measurement of $\Xi_b(6227)^-$ (see Fig. \ref{Fig2}), which also enforces the above conclusion.

Since two $2S$ $\Xi_b^\prime$ states are still missing in experiment, we provide the whole information of their strong decays in Table \ref{table4}. Obviously, the $\Lambda_b K$ and $\Xi_b\pi$ channels can be applied to reconstruct the signals of the predicted $\Xi^{\prime}_b(6365)$ and $\Xi^{\prime}_b(6376)$.

\begin{table}[htbp]
\caption{The partial and total decay widths of the $2S$ $\Xi_b$ and $\Xi_b^\prime$ states (in MeV). The forbidden decay channels are denoted by the symbol $\times$. If the threshold of a decay channel lies above the excited state, we denote the channel by $-$.} \label{table4}
\renewcommand\arraystretch{1.22}
\begin{tabular*}{85mm}{@{\extracolsep{\fill}}cccc}
\toprule[1pt]\toprule[1pt]
Decay modes                & $\Xi_b(6255)$  & $\Xi^{\prime}_b(6365)$  & $\Xi^{\prime}_b(6376)$ \\
\midrule[0.8pt]
 $\Lambda_bK$              & $\times$       & 15.6     & 16.1         \\
 $\Sigma_b(5815)K$         &   $-$          & 4.4      & 1.6         \\
 $\Sigma_b(5835)K$         &   $-$          & 0.8      & 3.3         \\
 $\Xi_b\pi$                & $\times$       & 16.0     & 16.4          \\
 $\Xi^{\prime}_b(5935)\pi$ & 6.3            & 7.2      & 1.9          \\
 $\Xi^{*}_b(5955)\pi$      & 10.2           & 3.4      & 9.4          \\
 $\Xi_b(6096)\pi$          & $\times$       & 5.9      & 1.7          \\
 $\Xi_b(6102)\pi$          & $\times$       & 2.5      & 7.4            \\
 \midrule[0.8pt]
  Total                    & 16.5           & 55.8    & 57.8      \\
\bottomrule[1pt]\bottomrule[1pt]
\end{tabular*}
\end{table}

\subsection{$1P$ states}

There are five $1P$ $\Xi_b^\prime$ states. For distinguishing them, we introduce the notation $\Xi^{\prime}_{bJ_{di}}(J^P)$, where $J_{di}$ is defined as the total angular momentum of the light diquark, i.e., $J_{di}\equiv\vec{S}_{di}+\vec{L}$. With our predicted masses as input, the partial widths of the strong decays of these five $\Xi_b^\prime$ states are obtained (see Table \ref{table5}). Although the $\Xi^{\prime}_{b0}$ state can decay into both $\Lambda_bK$ and $\Xi_b\pi$ channels, the partial decay width of the $\Xi_b\pi$ mode is much smaller than that of the $\Lambda_b K$ mode. This fact contradicts the experimental result of the $\Xi_b(6227)^-$, where the ratio of the partial decay widths of the $\Xi_b(6227)^-$ into $\Lambda_bK$ and $\Xi_b\pi$ is about 1~\cite{Aaij:2018yqz}.
Thus, the possibility of the $\Xi_b(6227)^-$ as the $\Xi^{\prime}_{b0}$ state can be excluded.

Two $\Xi^{\prime}_{b1}$ states with $J^P=1/2^-$ and $J^P=3/2^-$ cannot decay into $\Lambda_b K$ and $\Xi_b\pi$, which implies that the explanation of the $\Xi_b(6227)^-$ as the $\Xi^{\prime}_{b1}$ state with $J^P=1/2^-$ or $J^P=3/2^-$ can be also killed. We further illustrate that the main decay mode of the $\Xi^{\prime}_{b1}$ state with $J^P=1/2^-$ is $\Xi^\prime_b(5935)\pi$ while the $\Xi^{\prime}_{b1}$ state with $J^P=3/2^-$
mainly decays into $\Xi^*_b(5955)\pi$. This information is crucial to experimental exploration on these two missing bottom-strange baryons.

\begin{table}[htbp]
\caption{The partial and total decay widths of the $1P$ $\Xi_b^\prime$ states which contain a bad diquark (in MeV). To distinguish two states with the same $J^P$, we denote their different total angular momentum of the light diquark, $J_{di}$, as the subscript.} \label{table5}
\renewcommand\arraystretch{1.22}
\begin{tabular*}{86mm}{@{\extracolsep{\fill}}lccccc}
\toprule[1pt]\toprule[1pt]
Decay  &\multicolumn{2}{c}{$1/2^-$}  & \multicolumn{2}{c}{$3/2^-$}  & $5/2^-$  \\
\cline{2-3}\cline{4-5}\cline{6-6}
modes  & $\Xi^{\prime}_{b0}(6249)$  & $\Xi^{\prime}_{b1}(6239)$  & $\Xi^{\prime}_{b1}(6244)$ & $\Xi^{\prime}_{b2}(6213)$ & $\Xi^{\prime}_{b2}(6217)$   \\
\midrule[0.8pt]
 $\Lambda_bK$               &  9.1     & $\times$ & $\times$  & 10.2   &  11.0     \\
 $\Xi_b\pi$                 &  0.2     & $\times$ & $\times$  & 11.4   &  11.7    \\
 $\Xi^{\prime}_b(5935)\pi$  & $\times$ & 15.1     & 0.9       & 1.0    &  0.5      \\
 $\Xi^{*}_b(5955)\pi$       & $\times$ & 2.0      & 23.7      & 1.0    &  1.7        \\
 $\Xi_b(6096)\pi$           &  0.3     & 0.1      & 0.1       &  $-$   &  $-$   \\
 $\Xi_b(6102)\pi$           &  0.3     & $-$      & 0.1       &  $-$   &  $-$     \\
 \midrule[0.8pt]
  Theory                    & 9.9      &  17.2    & 24.8      & 23.6  &  24.9      \\
  Expt.~\cite{Aaij:2018yqz} &          &          &           & \multicolumn{2}{c}{$18.1\pm5.4\pm1.8$}        \\
\bottomrule[1pt]\bottomrule[1pt]
\end{tabular*}
\end{table}
%\end{widetext}

Indeed, our theoretical results support the possibility of the observed $\Xi_b(6227)^-$ as a $P$-wave $\Xi^{\prime}_{b2}$ state with $J^P=3/2^-$ or $J^P=5/2^-$ due to three pieces of evidence: 
\begin{enumerate}[(1)]
\item These two bottom-strange baryons have comparable partial widths for the decay modes $\Lambda_bK$ and $\Xi_b\pi$. And,
the ratio of the partial widths of $\Lambda_b K$ and $\Xi_b \pi$ for these two bottom-strange baryons is around 1, which is consistent with the measurement of $\Xi_b(6227)^-$ from LHCb~\cite{Aaij:2018yqz}.

\item The predicted total widths of these two states are also comparable with the experimental data.

\item The large partial widths of the $\Lambda_bK$ and $\Xi_b\pi$ decay channels can also explain why the $\Xi_b(6227)^-$ was first observed in these two decay modes~\cite{Aaij:2018yqz}.
\end{enumerate}

The results listed in Table \ref{table5} show the similarity of the decay behaviors of the $\Xi^{\prime}_{b2}$ states with $J^P=3/2^-$ and $J^P=5/2^-$. It results in the difficult situation of how to further distinguish these two assignments to the observed $\Xi_b(6227)^-$ according to the present experimental information.

In Ref.~\cite{Wang:2017kfr}, the decay widths obtained by a constituent quark model also supported the new $\Xi_b(6227)^-$ as a $P$-wave $\Xi^{\prime}_b$ candidate. However, some differences exist between our results and these presented in Ref.~\cite{Wang:2017kfr}. First, with the $LS$ coupling scheme, five $P$-wave $\Xi^{\prime}_b$ states can decay into both $\Lambda_b K$ and $\Xi_b \pi$ channels in Ref.~\cite{Wang:2017kfr}. Differently, only the $\Xi^{\prime}_{b0}(1/2^-)$, $\Xi^{\prime}_{b2}(3/2^-)$, and  $\Xi^{\prime}_{b2}(5/2^-)$ states can decay into $\Lambda_b K$ and $\Xi_b \pi$ in our $JJ$ coupling scheme. Second, as a $J^P=3/2^-$ or $J^P=5/2^-$ $\Xi^{\prime}_b$ state, the partial width ratio between the $\Lambda_bK$ and $\Xi_b\pi$ decay modes in Ref.~\cite{Wang:2017kfr} was predicted to be about 0.26$\sim$0.37, which is much smaller than our result (see Table \ref{table5}).

\begin{table}[htbp]
\caption{The partial and total decay widths of the $1P$ $\Xi_b$ states which contain a good diquark (in MeV).} \label{table6}
\renewcommand\arraystretch{1.3}
\begin{tabular*}{85mm}{@{\extracolsep{\fill}}cccc}
\toprule[1pt]\toprule[1pt]
Decay modes        & $\Xi^{\prime}_b(5935)\pi$  & $\Xi^{*}_b(5955)\pi$   & Total \\
\midrule[0.8pt]
 $\Xi_b(6096)$     & 4.2                        &    $1.0\times10^{-3}$  & 4.2        \\
 $\Xi_b(6102)$     & $1.0\times10^{-2}$         &     2.9                & 2.9        \\
\bottomrule[1pt]\bottomrule[1pt]
\end{tabular*}
\end{table}

Besides the five $1P$ $\Xi_b^\prime$ states discussed above, there are two $1P$ $\Xi_b$ states predicted in this work, i.e., the $\Xi_b(6096)$ and $\Xi_b(6102)$. If adopting the theoretical mass as input, the predicted $\Xi_b(6096)$ and $\Xi_b(6102)$ only decay into
$\Xi^{\prime}_b(5935)\pi$ and $\Xi^{*}_b(5955)\pi$ (see Table\ref{table6}). Here, the dominant decay mode of the $\Xi_b(6096)$ is $\Xi^{\prime}_b(5935)\pi$, while the $\Xi^{*}_b(5955)\pi$ has a dominant contribution to the total decay width of the predicted $\Xi_b(6102)$. The main conclusion obtained here for two $1P$ $\Xi_b$ states is consistent with that in Ref.~\cite{Wang:2017kfr}.

\section{Discussion and conclusion}\label{sec3}

Very recently, the LHCb discovered a new bottom-strange baryon state, the $\Xi_b(6227)^-$, by studying the $\Lambda_bK$ and $\Xi_b\pi$ invariant mass spectra~\cite{Aaij:2018yqz}. Stimulated by this new state, we carry out a phenomenological analysis of the $2S$ and $1P$ $bsq$ states by performing the mass spectrum analysis and the two-body OZI-allowed strong decay investigation. Finally, two assignments of the $\Xi_b(6227)^-$ become possible, i.e., the $\Xi_b(6227)^-$ can be explained as a $1P$ $\Xi^{\prime}_b$ state with either a $J^P=3/2^-$ or $J^P=5/2^-$ quantum number. Our study also shows that the decay property of the $P$-wave $\Xi^{\prime}_b$ state with $J^P=3/2^-$ is very similar to the state with $J^P=5/2^-$. So, distinguishing these two possible assignments becomes difficult only by the present experimental data.

Besides decoding the inner structure of the $\Xi_b(6227)^-$, we also predict the existence of its partners by illustrating their mass spectrum and decay behaviors. Surely, these results are helpful to find more excited bottom-strange baryons in a future experiment like the LHCb.

We also believe that the newly observed $\Xi_b(6227)^-$ is only a good start point when we are constructing the excited bottom baryon family. With the running of the LHCb in Run II, more and more data will be accumulated. In the near future, theorists and experimentalists should pay more attention to this interesting research issue.

\section*{Acknowledgement}

This project is supported by the National Natural Science Foundation of China under Grants No. 11305003, No. 11222547, No. 11175073, No. 11647301, No. U1204115 and No. 11475111. Xiang Liu is also supported by the National  Program for Support of Top-Notch Young Professionals.

%\end{widetext}

\end{document}